\documentstyle[manuscript,aps]{revtex}

\begin{document}
\preprint{UM-P-96/42 }

\draft

\title{THE CP-ODD NUCLEON INTERACTION AND THE VALUE OF T-VIOLATION IN
NUCLEI\footnote{Talk at Trento Workshop, October 1995} }
\author{Vladimir P. Gudkov}
\address{School of Physics\\
University of Melbourne \\
Parkville, Vic. 3052 Australia}

\maketitle
\begin{abstract}
The relations between the value of T- and P-violating correlations in
neutron scattering and different models of CP violation are discussed. It is
shown that a specific structure of CP-odd nucleon interactions gives the
possibility to obtain the essential information about CP-odd interaction at
the quark-gluon level from nuclear experimental data. The up-to-date
estimations for CP-violating nucleon coupling constants show that CP
violation in neutron scattering is sensitive to many models of CP violation.
\end{abstract}

\pacs{ }

\newpage   

\section{INTRODUCTION}

From the first experimental discovery of CP violation in the $K^0$-meson
decays in 1964 still there are no other experimental confirmation of CP
violation in other systems in spite of many attempts in high energy, nuclear
and atomic physics. The problem of CP violation is closely related to the
time reversal invariance because, according to the CPT theorem, the
violation of CP-invariance means the violation of time reversal invariance
(T violation). Moreover, in low energy physics T violation is the only one
possibility of manifestation of CP violation. It should be noted, that T
violation has never been observed and, therefore, it is a subject of a
special interest regardless of the CP violation itself. The problem of T
violation has been a subject of an experimental investigation in nuclear
physics for several tens years. In general, there are two different
possibilities for T- violation: due to CP-odd and P-odd (or T- and
P-violating) interactions or due to T-violating P-even interactions. The
first case is related to the one observed in $K$- meson decays, and the
second one -- to phenomenological interactions which may not exist at all.
Here we consider the first case. For the simplicity of terminology and in
accordance to the common believe in the CPT theorem we use terms CP
violation and T violation as synonyms. This does not restrict an idea about
possible source of T violation and all conclusions are independent on a
nature of T violation with one exception: calculations of CP violating
nucleon coupling constants will be done CP- and P-violating interactions.

Investigations of T violation in resonance neutron scattering give the
possibility to improve drastically the current situation in searching for T
violation (see, e.g. paper\cite{nucl} and references therein). As it has
been shown\cite{bgtf}, the main advantage in searching for CP violation in
neutron scattering is a large expected value for CP-odd effects due to
nuclear enhancement factors. The expected large value of nuclear effects
might be a big disadvantage, also. A good illustration for such a situation
is the well known large parity violation in neutron induced reactions where
the complexity of the system leads to impossibility for a theoretical
description of P-odd effects in nuclei in terms of  quark-gluon weak
interactions. Therefore, P-odd effects are used to study the nuclear
structure rather than the weak interactions  which can be studied better in
high energy physics.

Such a situation can not be satisfactory  for the CP violation. Since the
source of CP violation is still unknown, the primary aim is to find any
manifestation of CP violation and to measure its intensity in terms of
quark-gluon coupling constants to choose the appropriate mechanism of CP
violation. For this purpose one must go from the experimental data in
neutron scattering up to quark-gluon coupling constants and to solve the
problem of theoretical description of the nuclear experimental data in terms
of parameters of CP-violating models. The problem of a reliable
interpretation of the experimental data is even more important for the
improving of the current experimental restrictions on the CP-violating
coupling constants if CP violating effects will be not detected in
experiments.

To calculate various CP violating effects in nuclei we need go through the
different levels of theoretical models. At first, one needs to obtain the
effective low energy CP violating Lagrangian at the quark level for the
particular model of CP violation. The second step is the calculation of the
CP-violating nucleon-nucleon interaction using the obtained low energy
Lagrangian. The third step is the calculation of nuclear CP-violating
effects in a nuclear model using calculated CP-violating nucleon couplings.
Each step of these calculations is a model dependent, in general.

For example, in order to describe the standard P-odd and CP-even
nucleon-nucleon interaction it is necessary to calculate at least six
different meson-nucleon coupling constants\cite{ddh,adelb,dub}. The masses,
spin and isospin properties of these mesons are different from each other.
Therefore, the obtained P-odd nucleon potential is rather complex and, being
used in nuclear models, it leads to big uncertainties in calculations of
P-odd nuclear effects.

It will be shown that the problem of interpretation of the CP-violating
experimental data in neutron scattering might be solved successfully and one
can get rid of the model dependencies at all levels of the calculation.

Let us recall the main results for the description of CP-violating effects
in neutron scattering related to the T-odd and P-odd correlation $({\vec
\sigma [\vec k\times \vec I])}$, where $\vec \sigma $ and ${\vec I}$ are
neutron and target spins, and ${\vec k}$ is the neutron momentum. This
correlation leads \cite{kabirf,stodtf} to the difference of the total cross
sections for the transmission of neutrons, polarized parallel and
antiparallel to the axis ${\ [\vec k\times \vec I]}$, through the polarized
target

\begin{equation}
\label{sicp}\Delta \sigma _{CP}={\frac{{4\pi }}k}{\it Im}(f_{\uparrow
}^p-f_{\downarrow }^p), 
\end{equation}
and to the neutron spin rotation angle $\chi $ around the axis ${[\vec
k\times \vec I]}$

\begin{equation}
\label{hicp}{\frac{{d\chi }}{{dz}}}={\frac{{2\pi N}}k}{\it Re}(f_{\uparrow
}^p-f_{\downarrow }^p). 
\end{equation}
Here $f_{\uparrow ,\downarrow }^p$ are the zero angle scattering amplitudes
on the polarized nuclei for neutrons polarized parallel and antiparallel to
the ${[\vec k\times \vec I]}$ axis, respectively, $z$ is the target length,
and $N$ is the density of nuclei in the target.

It was shown \cite{bgtf,bgn} that T-violating parameters $\Delta \sigma
_{CP} $ and ${d\chi }/{dz}$ look like the P-violating ones caused by the
T-even P-odd correlation ${(\vec \sigma \vec k)}$ and have the analogous
enhancement factors which lead to their increase by a factor of $10^5-10^6$.
Their expressions in the two resonances approximation are

\begin{equation}
\label{dtot}\Delta\sigma_{CP} = - {\frac{{2\pi G_J}}{k^2}}{\frac{{%
w(\Gamma^n_s\Gamma^n_p(S))^{\frac{1}{2}}}}{{[s][p]}}}[(E - E_s)\Gamma_p + (E
- E_p)\Gamma_s], 
\end{equation}

\begin{equation}
\label{dfdz}{\frac{{d\chi }}{{dz}}}={\frac{{4\pi NG_J}}{k^2}}{\frac{{%
w(\Gamma _s^n\Gamma _p^n(S))^{\frac 12}}}{{[s][p]}}}[(E-E_s)(E-E_p)-{\ \frac
14}\Gamma _s\Gamma _p],
\end{equation}
where $[s,p]=(E-E_{s,p})^2+{{\Gamma _{s,p}^2}/4}$ and $E_{s,p}$, $\Gamma
_{s,p}$ and $\Gamma _{s,p}^n$ are the energy, total and neutron widths of
the $s$- and $p$-wave compound resonances, $w$ is imaginary (T-non-invariant
part) of the P-odd matrix element between these resonances, ${G_J\ }$is a
spin function dependent on the spin of compound system $J$ andthe channel
spin $S=I\pm 1/2$.

From eq.(\ref{dtot}) and from the corresponding expression\cite{bgn} for
P-violating difference of the total cross sections for the transmission of
neutrons with opposite helicities through an unpolarized target

\begin{equation}
\label{ptot}\Delta \sigma _P\sim {\frac{2{\pi }}{k^2}}{\frac{v{(\Gamma
_s^n\Gamma _p^n)^{\frac 12}}}{{[s][p]}}}[(E-E_s)\Gamma _p+(E-E_p)\Gamma _s]
\end{equation}
one can obtain the relation between the values of the P-odd and the CP-odd
effects\cite{gcp}

\begin{equation}
\label{svyaz}\Delta \sigma _{CP}=\kappa (J){\frac w{{v}}}\Delta \sigma _P, 
\end{equation}
where $v$ is real part of the weak matrix element between the $s$- and $p$%
-resonance states. The parameter $\kappa (J)$ is

\begin{eqnarray}  
\kappa (I+1/2) = && -{3\over{2^{3/2}}}\left( {{2I+1}\over{2I+3}}\right) ^{3/2}   
\left( {3\over{\sqrt{2I+3}}}\gamma - \sqrt{I} \right) ^{-1}, \\    
 \kappa (I-1/2) = && -{3\over{2^{3/2}}}\left( {{2I+1}\over{2I-1}}\right) \left(   
{I\over{I+1}}\right) ^{1/2} \left( -{{I-1}\over{\sqrt{2I-1}}}{1\over{\gamma}} + \sqrt{I+1}   
\right) ^{-1}.  \label{svsp}  
 \end{eqnarray}   

Here $\gamma =[\Gamma _p^n(I+1/2)/\Gamma _p^n(I-1/2)]^{1/2}$ is the ratio of
the neutron width amplitudes for the different channel spins. In general,
the parameter $\gamma $ may be obtained from the angular correlation
measurements in neutron induced reactions.

The P-odd parameter $\Delta \sigma _P$ has been measured in many
experiments. Its relative value (compared with the neutron total cross
section) in the vicinity of p-wave resonances has a huge magnitude for weak
processes, up to $10^{-1}$(see, e.g. ref.\cite{alfim} and references
therein).

From eq.(\ref{svyaz}) one can see\ that the measurement of the CP-odd and
P-odd effects at the same $p$-wave compound resonance (when the values reach
their maximum) leads to the possibility of extracting the ratio

\begin{equation}
\label{srl}<\lambda >={\frac w{{v}}}. 
\end{equation}

Due to the large value of the P-odd parameter $\Delta \sigma _P$ in the
vicinity of p-wave compound resonance, there is the possibility to measure
the CP-odd parameter $<\lambda >$ at a level up to $10^{-4}$(see, e.g. refs.%
\cite{bow,masuda}). 

However, the parameter $<\lambda >$ is the ratio of the CP-odd to the P-odd
matrix elements between s- and p-wave compound resonances, but not the ratio
of nucleon (or quark) coupling constants. The structure of compound
resonances is very complicated and is usually described by statistical
methods. Therefore, we have a large experimental effect and can not obtain
an information from the experimental result due to enormous difficulties in
the theoretical descriptions. The reasons for the enhancement and for the
theoretical difficulties are the same - the complexity of nuclear compound
states.

To avoid such deadlock we explore an approach where only ratios of CP-odd to
P-odd parameters are calculated at all levels of a hierarchy of the models.
If the possible structures of the CP-odd and P-odd interactions are known at
each level (nuclear interactions, nucleon interactions and quark-gluon
interactions) the calculations of the ratio for CP-odd to P-odd parameters
give the opportunity to eliminate many model dependent features. The
important point is that the experimental values of the P-odd parameters are
known and their theoretical values have been calculated. It will be shown
that the structure of CP-odd nucleon interactions is simpler than the
structure of P-odd ones. This fact gives an additional simplification in
obtaining of the CP-odd coupling constants from the nuclear experimental
data.

Let us start from the consideration of the ratio of the compound nuclear
matrix elements $<\lambda >$.

\section{The relation between nuclear matrix elements and nucleon coupling
constants}

To estimate the parameter $<\lambda >$ we can use a simple model of the one
particle interaction. Then the one particle potentials for P-violating \cite
{mitch} and CP-violating \cite{sfkh} interactions are:

\begin{mathletters}

\label{eq:pot}
\begin{equation}  
V_P = {G_F\over{8^{1/2}M}} \{ ({\vec \sigma \vec p}),\rho (\vec r)\} _+ , \label{eq:a}
\end{equation}
\begin{equation}     
V_{CP} = {{iG_F\lambda }\over{8^{1/2}M}} \{ ({\vec \sigma \vec p}),\rho (\vec r)\} _- ,
\label{eq:b}   
\end{equation}   

\end{mathletters}

where $G_F$ is the weak interaction Fermi constant, $M$ is the proton mass, $%
\rho ({\vec r)}$ is the nucleon density, ${\vec p}$ is the momentum of the
valence nucleon and $\lambda $ is the ratio of CP-violating to P-violating
nucleon - nucleon coupling constants.

Now one obtains from eqs.(\ref{srl}) \ and (\ref{eq:pot}) that

\begin{equation}
\label{svl}<\lambda > = {\frac{\lambda}{{1+2\xi}}}, 
\end{equation}

where

\begin{equation}
\label{ksi}\xi = {{\frac{{\langle {\phi_p}|\rho ({\vec \sigma \vec p})| {%
\phi_s}\rangle }}{{\langle {\phi_p}|{(\vec \sigma \vec p)}\rho | {\phi_s}
\rangle}}}}. 
\end{equation}

Here $\phi_{s,p}$ are the $s,p-$resonance wave functions of the compound
nucleus.

Let us consider the matrix elements in eq.(\ref{ksi}). The operator identity 
$2{\vec p}=iM[H,{\vec r}]$ leads to the value of the numerator \cite{gcp}

\begin{equation}
\label{num}{\langle {\phi_p}|\rho ({\vec \sigma \vec p})| {\phi_s}\rangle }
\simeq {\frac{{i\overline{\rho}M}}{2}}D_{sp}{\langle {\phi_p}|({\vec \sigma
\vec r})|{\phi_s} \rangle }. 
\end{equation}

Here $H$ is the single particle nuclear Hamiltonian, $D_{sp}$ is the average
single particle level spacing, and $\overline{\rho}$ is the average value of
the nuclear density. The denominator of eq.(\ref{ksi}) \ is

\begin{eqnarray}  
\langle {\phi_p}|{(\vec \sigma \vec p)}\rho | {\phi_s}\rangle  & =  - \langle {\phi_p}|{(\vec \sigma   
\vec r)}{1\over{r}}{{\partial \rho} \over{\partial r}} | {\phi_s}\rangle \nonumber \\    
& =  {{2i\rho}\over{R^2}}\langle {\phi_p}|{(\vec \sigma \vec r)} | {\phi_s}\rangle ,    \label{equ1}  
\end{eqnarray}   

where $R$ is the nuclear radius.

Inserting eqs.(\ref{num}) \ and (\ref{equ1}) \ into eq.(\ref{ksi}) \ we
obtain

\begin{equation}
\label{ksies}\xi ={\frac 1{{4}}}MD_{sp}R^2={\frac 1{{4}}}\pi (KR), 
\end{equation}

where\ we used the estimate of $D_{sp}$ for the square-well potential case 
\cite{bm}:

\begin{equation}
D_{sp}={\frac 1{{MR^2}}}\pi KR,
\end{equation}
and $K$ is the nucleon momentum in the nucleus. Eq.(\ref{ksies}) \ gives the
numerical values for  $\xi \in (1\div 7)$. Therefore, we can conclude that
the values of the matrix elements in eq.(\ref{ksi}) \ are of the same order
of magnitude and that, consequently, the values of $<\lambda >$ and $\lambda 
$ are of the same order of magnitude, as well. In other words, there are no
large suppression factors in the relation between of $<\lambda >$ and $%
\lambda $ parammeters and, therefore, models of CP violation might lead to
measurable values of CP-violating effects in neutron scattering. Also, we
arrive at the possibility to distinguish between the models of CP violation,
which have different CP-odd nucleon-nucleon coupling constants. It should be
noted, that when the experimental data will be available, the ratio $%
<\lambda >/\lambda $ should be calculated more accurately for each
particular nucleus using a realistic approximation for the nuclear density
and wave functions.

If we consider the eq.(\ref{svl}) not as a rough approximation for the given
above estimation of the ratio of the matrix elements but rather more
seriously, we will come to the conclusion that the CP-odd matrix elements
have a regular suppression factor $\left( 1+2\xi \right) $ comparing to the
P-odd ones. This result is in a good agreement with the detailed numerical
studies (see, e.g. refs.\cite{vogel,towner}). Furthermore, neglecting the
first term in the denominator of eq.(\ref{svl}) one has got the parametrical
suppression factor \cite{khripl} for CP-odd matrix elements

\begin{equation}
\frac{<\lambda >}\lambda \simeq \frac 2\pi K^{-1}r_0^{-1}A^{-1/3}\sim
A^{-1/3}, 
\end{equation}
where $R=r_0A^{1/3}$ and $A$ is the atomic number. The existence of this
possible suppression factor can be explained by the fact that \cite{desp5}
the CP-odd nuclear potential has the well-defined surface character and,
therefore, it is proportional to the size of nuclear surface ($4\pi R^2\sim
A^{2/3}$) , but the P-odd nuclear potential has the volume character and it
is proportional to the nuclear volume ($\frac{4\pi }3R^3\sim A$).

\section{The\ CP-odd nucleon coupling constants}

To obtain a relation between the experimental data and the possible models
of CP violation it is necessary to calculate the CP-odd nucleon coupling
constants in these models. It is well known from the experience of
calculations of P-odd nuclear interactions that this is a very difficult and
sometimes an ambiguous procedure. To calculate P-odd nucleon interactions
the following common problems have to be sold:

\begin{itemize}
\item  to choose a model for description of nucleon-nucleon interactions;

\item  to calculate an effective symmetry violating Lagrangian, taking into
account quark-gluon interactions at short distances, and to renormalize the
Lagrangian to the nucleon scale;

\item  to calculate meson-nucleon P-odd interactions using hadron models.
\end{itemize}

The one-boson exchange model is usually used to describe nucleon-nucleon
interactions. The $\pi -$, $\rho -$, and $\omega -$meson exchanges are
taking into account. The effective weak Lagrangian is calculated on the base
of the standard model with QCD gluonic corrections at the short distances.
The procedure of renormalization to the large distances (the nucleon scale)
leads to the expression of the Lagrangian as a sum of terms with an accuracy
up to $O(\alpha _s\ln (M_W^2/\mu ^2)),$ where $\alpha _s$ is a strong
coupling constant, $M_W$ is the mass of $W$-boson and $\mu $ is the
parameter of the hadronic scale. It should be noted, that the QCD
perturbation theory is not applicable at the hadronic scale ($\mu \sim 1GeV$%
), therefore, the renormalization procedure up to the level $\mu $ is not
correct and might be a source of uncertainties in the further calculations.
However, the source causing most uncertainties is the last step: the
calculation of P-odd meson - nucleon coupling constants using hadron models. 

Let us shortly overview some of the existing approaches for these
calculations which are based on: quark models, topological soliton models,
the chiral perturbation theory and QCD sum rules.

In the traditional approach using the quark model (see, for example \cite
{ddh,adelb,dub} ) the $M$-meson nucleon weak matrix element $\left\langle
MN^{\prime }\left| {\cal L}_{PV}\right| N\right\rangle $ might be
represented as a sum of two parts:

\begin{equation}
h_M=\left\langle MN^{\prime }\left| {\cal L}_{PV}\right| N\right\rangle
=h_M^F+h_{M,}^{NF} 
\end{equation}

where $h_M^F$ is so called ''factorized'' or calculable in the factorization
approach part and $h_M^{NF}$ is the '' non-factorized'' part; ${\cal L}_{PV}$
is the effective weak Lagrangian. The factorized part of the matrix element
for the case of $\pi $-mesons 
\begin{equation}
h_\pi ^F\sim \left\langle \pi ^{-}\left| \overline{d}\gamma _5u\right|
0\right\rangle \cdot \left\langle p\left| \overline{u}d\right|
n\right\rangle , 
\end{equation}
can be calculated using equations of motion for the quarks

\begin{equation}
\left\langle \pi ^{-}\left| \overline{d}\gamma _5u\right| 0\right\rangle = 
\frac{f_\pi \cdot m_\pi ^2}{(m_u+m_d)}, 
\end{equation}

\begin{equation}
\left\langle p\left| \overline{u}d\right| n\right\rangle =\frac{M_\Xi
-3M_\Lambda +2M_p}{m_u-m_s}. 
\end{equation}

For the case of vector mesons ($\rho $, $\omega $), the vector meson
dominance hypothesis is used for calculating of the vector current matrix
element in the factorized part

\begin{equation}
h_\rho ^F\sim \left\langle \rho \left| V_\mu \right| 0\right\rangle \cdot
\left\langle N^{\prime }\left| A^\mu \right| N\right\rangle . 
\end{equation}

The non-factorized part $h_M^{NF}$ can be calculated only numerically (e.g.,
in a quark bag model), therefore, it is the main source of uncertainties in
these calculations. These uncertainties lead to a rather large range for the
value of the weak $\pi $-meson coupling\cite{ddh,adelb,dub}: $h_\pi \in $ $%
(1\div 5)\cdot 10^{-7\text{ }}$.

The topological soliton model approach has been used in last years\cite
{kaiz,ulf} to calculate $\pi $-meson weak coupling constants. Its advantage
is the possibility of simultaneous calculations of the strong and weak
interaction regions. However, it does not predict all nucleon properties
and, therefore, its accuracy and reliability does not look satisfactory
enough. In this approach nucleons are considered as solitons of a non-linear
meson theory. The Hamiltonian of weak interaction is rewriting in terms of
currents constructed from meson fields, which make up the soliton. The meson
field is represented as a sum of two components: one, that makes up the
soliton and another, which is a small pionic fluctuation. The linear in the
pionic fluctuation terms correspond to $\pi $-meson nucleon interactions and
the quantalization of the respective operator gives the coupling constant in
the following form\cite{ulf}

\begin{equation}
h_\pi =8\pi \frac{G_F\sin {}^2\Theta _W}{\Theta \cdot f_\pi }%
\int\limits_0^\infty r^2[I_0(r)V_0(r)-\frac 23I_1(r)V_1(r)]dr, 
\end{equation}

where $\Theta _{W\text{ }}$ is the Weinberg angle, $\Theta $ is the moment
of inertia of the spinning soliton, $I_{0,1}(r)$ and $V_{0,1}(r)$ are the
radial functions associated with the time (space) components of the
isoscalar (isovector) soliton current, respectively. This expression gives
the numerical value $h_\pi \sim 0.25\cdot 10^{-7}$.

The chiral perturbation theory approach to weak nucleon interactions\cite
{kaplan} gives the low energy weak meson-nucleon Hamiltonian which includes
large $\pi \pi NN$ and $\gamma \pi NN$ couplings. However, in this approach
the meson-nucleon coupling constants can be calculated only numerically,
using lattice methods. The dimensional analysis used in ref.\cite{kaplan}
gives the following $\pi $-meson nucleon coupling constant

\begin{equation}
h_\pi \simeq \left( \frac{\Lambda _\chi }{f_\pi }\right) \cdot \frac{G_F}{
\sqrt{2}}\cdot f_\pi ^2\simeq 5\cdot 10^{-7}. 
\end{equation}

It should be noted, that in the chiral approach the strange quarks give a
large contribution to the weak coupling constants (see, paper\cite{kaplan}
and references therein).

The QCD sum rules applied for the calculation of  weak $\pi $-meson coupling
constants\cite{khat,henl} give a rather different results: $h_\pi \sim
5\cdot 10^{-7}$ in the ref.\cite{khat} and $h_\pi \sim 0.3\cdot 10^{-7}$ in
the ref.\cite{henl}. As it has been stated in the last paper, the smaller
result is obtained due to the cancellation between perturbative and
non-perturbative QCD modifications of the weak process.

From the given cosideration of different approaches for the calculation of
P-odd meson-nucleon coupling constants we can see that  this a rather
difficult problem which is still far from the final resolution. These
difficulties are common for the calculations of both the P-odd and the
CP-odd coupling constants, but in the case of CP violation an additional 
complexities exist: there are many various possible models for CP violation.
Fortunately, the large number of these models leads to a few different
structures for CP-violating low energy Lagrangians. Moreover,it will be
shown that many of the existing problems for calculations of P-odd
interactions can be simplified or even eliminated for calculations of CP-odd
interactions if we calculate not the coupling constants themselves but the
ratio of them to the P-odd coupling constants with the same (or, almost the
same) structure of the effective Lagrangians. This approach gives a real
advantage in eliminating some model dependant uncertainties arising from the
strong interactions at the quark-gluon and nucleon levels. The next
simplification is the fact\cite{ghm}, that to calculate the CP-odd nucleon
interactions it is enough to to take into account only $\pi $-meson
contributions. Therefore, we can use only the one $\pi $-meson - nucleon
interaction to calculate all CP violating effects in nuclei. This is a very
important point to be considered here (see ref.\cite{ghm}).

For the low energy region all CP violating models can be grouped into four
classes according to the sources of CP violation on the quark-gluon level:

\begin{itemize}
\item[a.]  Complex quark mass matrices. In the mass eigenstate basis, there
will be CP violation in the charged current due to exchange of gauge
particles. One of the best known example is the Kobayashi-Maskawa model\cite
{km}.

\item[b.]  Complex mixing angles for gauge bosons. An example is the
left-right symmetric model\cite{lr}.

\item[c.]  Complex vacuum expectation values of Higgs bosons, for example
the Weinberg model\cite{wein}.

\item[d.]  CP-odd pure gluonic interaction, and as the $\theta $-term in QCD%
\cite{theta}.
\end{itemize}

In a specific process, some or all of these CP violating sources contribute
and corresponding effective Lagrangians include CP-odd pure quark,
quark-gluon and pure gluonic operators. The pure quark operators appear in
the form current $\times $ current due to gauge boson exchange, or
pseudo-scalar $\times $ scalar structure due to scalar boson exchange. The
most important feature of these Lagrangians is the presence of the
right-current $\times $ left-current or the pseudo-scalar $\times $ scalar
structures. These operators have enhanced contributions to the CP-odd
pseudo-scalar meson-nucleon couplings. This is a principal difference
compared to the structure of the P-odd and CP-even effective Lagrangian
which leads to the enhancement of pseudo-scalar $\times $ scalar
contribution and, as a consequence, to decreasing the number of
meson-nucleon coupling to just one $\pi $-meson interaction with nucleon.

Let us consider the low energy effective Lagrangian involving only u and d
quarks (operators up to dimension six are considered). Exchanging gauge
bosons at the tree level in the a- and b-type of models will produce the
following structure of the Lagrangian

\begin{eqnarray}
{\cal L}&\sim& L\times L + L\times R + R\times L + R\times R\nonumber\\
&=& C_{LL}O_{LL} + C_{LR}O_{LR} + C_{RL}O_{RL} + C_{RR}O_{RR}\;,
\end{eqnarray}

where $O_{LL}=\bar u_L\gamma _\mu d_L\bar d_L\gamma _\mu u_L$ and other
operators are defined in a similar way. Note that only $L\times R$ and $%
R\times L$ have CP violating interaction.

At the tree level the c-type of models will lead to the CP violating
effective Lagrangian

\begin{equation}
\label{eq1}{\cal L}\sim S\times P=C_{SP}\bar q_1q_2\bar q_3\gamma
_5q_4+h.c.\;, 
\end{equation}
where $q_i$ can be $u$ and $d$ quarks depending whether a charged or neutral
scalar is exchanged to produce the effective Lagrangian.

The $L\times R$ term also contains a term proportional to $S\times P$. This
can be seen by making a Fierz transformation on $O_{LR}$. We have

\begin{equation}
O^F_{LR} = -2[ {\frac{1}{3}}\bar u_L u_R\bar d_R d_L + {\frac{1}{2}} \bar
u_L \lambda^a u_R \bar d_R \lambda^a d_L] + h.c. \;. 
\end{equation}

We can now calculate the CP-odd meson-nucleon coupling constants for $\pi $-
and $\rho $- mesons from the $L\times R$ term. Using the factorization
approximation and the vector meson dominance hypothesis, we have

\begin{eqnarray}
\bar g _{\pi^- NN}&\approx& <\pi^- p|L\times R|n>_{CP} = i{ImC_{LR}\over 2}
<\pi^-|\bar d \gamma_\mu\gamma_5 u|0><p|\bar u \gamma_\mu d|n>\nonumber\\
&=& i{Im C_{LR}\over 2}{m_d^2 -m_u^2 \over m^2_\pi} <\pi^-|
\bar d \gamma_5 u |0><p|\bar u d |n>\;,\nonumber\\
\bar g_{\rho NN} &\approx& <\rho^- p|L\times R|n>_{CP} = {Im C_{LR}\over 2}
{m^2_\rho\over f_\rho}g_A\;. \label{eqfo}
\end{eqnarray}

where $m_\rho $ and $f_\rho $ are the mass and strong form factor for $\rho $%
-meson with $f_\rho ^2/4\pi \approx 2$, $g_A$ is the nucleon axial form
factor, $m_u$ and $m_d$ are masses of $u$- and $d$- quarks. Using the Fierz
transformed operator $O_{LR}^F$ and the factorization approximation, we
obtain

\begin{eqnarray}
\bar g _{\pi^0 NN} &\approx& {1\over 3} Im C_{LR}
(<\pi^0|\bar d \gamma_5d|0><N|\bar  u u|N> - <\pi^0|\bar u \gamma_5 u|0>
<N|\bar d d|N>)\;.
\end{eqnarray}

From the above equation we clearly see that there is a suppression factor $%
(m_d^2-m_u^2)/m_\pi ^2$ for $\bar g_{\pi ^{-}NN}$ compared with $\bar g_{\pi
^0NN}$.

To compare the contributions from the $\pi $ and $\rho $ meson exchanges to
the CP-odd nucleon potential we remind that for the standard P-odd and
CP-even interaction $L\times L$, we have the same structure of the
corresponding coupling constants

\begin{eqnarray}
g^p_{\pi^- NN} &\approx& {C_{LL}\over 2}{{m^2_d - m^2_u}\over m^2_\pi}
<\pi^-|\bar d \gamma_5 u|0> <p|\bar u d |n>\;.\nonumber\\
g^p_{\rho NN} &\approx& {C_{LL}\over 2} {m^2_\rho \over f_\rho}g_A\;. \label{eqsix}
\end{eqnarray}

It is well know that for the P-odd and CP-even nucleon potential the
contributions from the $\pi $ and $\rho $ mesons have the same order of
magnitude if the relative strength of the couplings is given by eq.(\ref
{eqsix})\cite{ddh}. It is expected that the same thing should happen for the
CP-odd nucleon potential. Then from eq.(\ref{eqfo}) we can see that the
contributions from the $\rho $ and $\pi ^{-}$ meson exchanges to the CP-odd
potential will have the same order of magnitude. Therefore, we conclude that
the dominant contribution to the CP-odd nucleon potential is from the $\pi ^0
$ meson exchange.

The similar results have been obtained\cite{ghm} for the c-type of models.
In this case, the $\rho $-meson nucleon coupling will be much smaller than $%
\pi $-meson nucleon couplings for the same reason as given above. However,
unlike the situation in the a- and b- type of models where the $\pi ^0$
meson-nucleon coupling is much larger than the $\pi ^{-}$ meson-nucleon
coupling, the charged and neutral pion-nucleon coupling can be of the same
order of magnitude. Therefore $\pi ^{\pm }$ and $\pi ^0$ exchange can all
make significant contributions to the CP-odd nucleon potential. The reason
for this enhancement on $\bar g_{\pi NN}$ is due to the large contribution
of the pseudo-scalar and scalar quark densities in the local approximation.
A similar enhancement factor for the strange quark current has been found in
penguin induced K-meson decays\cite{zvsh}.

CP-odd pure gluonic operators ($J^{PC}=0^{-+}$) can be generated in many
models\cite{mor,wingl}, particularly in the c- and d-type of models. It is
interesting to note, that because of the pseudo-scalar nature of the
operators, the pseudo-scalar meson-nucleon coupling constants are much
bigger than the vector meson-nucleon coupling constants, just as they are
for the pure quark operators. The estimation\cite{ghm} of the ratio of the
coupling constants for pseudo-scalar and vector mesons leads to the
conclusion that for gluonic CP-odd operators the coupling constant of the
pseudo-scalar meson to nucleon is larger by about one order of magnitude
than the vector meson-nucleon coupling constant. The same result is valid
for the lowest order CP-odd quark-gluon operator (the colour-electric dipole
moment) $\hat O=\bar q\sigma _{\mu \nu }\gamma _5({\lambda ^a/2)}qG^{a\mu
\nu }$. 

Now we can see that for all types of CP-violating models the contributions
to the CP-violating nucleon-nucleon interaction from pseudo-scalar mesons
are larger than the contributions from vector meson by about one order of
magnitude. Therefore, the dominant CP violating nucleon-nucleon interaction
in the one meson exchange approximation is from the $\pi $-meson exchange
and to calculate CP-odd effects in nuclei with a reasonable accuracy, we
need only consider pseudo-scalar meson exchange. (The dominant CP-odd $\pi $%
-meson contribution  has been confirmed by numerical calculations in paper%
\cite{towh}.) The situation here is quite different from the P-odd and
CP-even nucleon potential, where the $\pi $, $\rho $ and $\omega $ all
contribute significantly.

\section{Estimations of CP-odd nucleon interactions for different models}

Let us estimate the parameter $\lambda $ (the ration of CP-odd to P-odd
nucleon coupling constants) for some models of CP violation. We will keep to
the model classification given in the previous section.

\subsection{Models of class (a).}

The well known model of CP violation due to complex quark mass matrix is the
standard Kobayashi-Maskawa (KM) model\cite{km}. It gives negligible
contribution to the nucleon CP-odd interaction\cite{dhm,kky}:

$$
\lambda _{KM}\leq 10^{-10}. 
$$

Two other models with the similar source of CP violation are left- right\cite
{lr} and horizontal\cite{hori} models. The corresponding parameters $\lambda
_{LR}^q$ and $\lambda _H$ have been calculated in paper\cite{bgcon}. The
comparison of effective low energy Lagrangians for that models with the
Lagrangian for the Kobayashi-Maskawa model leads to the following
expressions:

\begin{equation}
\label{lqrl}\lambda _{LR}^q\sim \lambda _{KM}\left( {\frac{{M_L}}{{M_R}}}%
\right) ^2{\frac{{\sin {(\delta _2-\delta _1)}}}{{c_2s_2s_3\sin {\delta }}},}
\end{equation}

\begin{equation}
\label{lhor}\lambda _H\sim \lambda _{KM}{\frac{{4G_H\sin {\phi }}}{G{%
c_2s_1s_2s_3\sin {\delta }}}.} 
\end{equation}

Here $M_{L,R}$ are left(right)-handed gauge boson masses; $\delta $ is a
CP-odd phase in the Kobayashi-Maskawa model and the $\delta _{1,2}$ are
corresponding phases in the left- right model; $c_i=\cos {\theta _i}%
,s_i=\sin {\theta _i}$, where the $\theta _i$ are KM matrix mixing angles; $%
G_H$ and $\phi $ are the strength and CP-odd phase of the horizontal
interaction.

From eq.(\ref{lqrl}) it is obvious that the left-right model contribution in
the given scenario (Class (a)) is very small $\lambda _{LR}^q\leq \lambda
_{KM}$. Accepting the value for the strength of the horizontal interaction 
\cite{hori} ($10^{-16}GeV^{-2}\leq G_H\leq 10^{-11}GeV^{-2}$), we have got
the same conclusion for the horizontal model contribution: $\lambda _H\leq
\lambda _{KM}$. However, due to many uncertainties in the model, the case
with the parameter $\lambda _H\leq 10^{-4}$ cannot be ruled out, too (see,
e.g. ref.\cite{herc}).

It should be noted that the above estimations have been done for the
standard case of three quark generations. However, in the presence of the
fourth heavy quark generation the situation may be changed drastically. For
example, in the Kobayashi-Maskawa model with four quark generations
radiative electroweak corrections might lead to CP violation by several
orders of magnitude larger than for the standard three generations case
(see, e.g. calculations of the neutron electric dipole moment in ref.\cite
{forkm}).

\subsection{Models of class (b).}

The calculation of the CP violation due to complex mixing angles for gauge
bosons in the left-right symmetric model gives \cite{bgcon}

\begin{equation}
\label{lwrl}\lambda _{LR}^W\sim 10\sin {\zeta }\sin {\alpha }\simeq {\frac{{%
2\epsilon }}{{43}}}{\frac{{m_s}}{{m_c}}}, 
\end{equation}

\begin{center}
where $\zeta $ and $\alpha $ are the CP-even and the CP-odd mixing phases of
gauge bosons, $\epsilon $ is a CP-odd $K$-meson decay parameter, $m_s$ and $%
m_c$ are masses of $s$- and $c$-quarks 
\footnote{I appreciate Dr. P. Herczeg for the comment that the corresponding expression for  $\lambda _{LR}^W$ in ref.\cite{bgcon} has a superfluous factor $(M_L / M_R)^2$.}
. Using the experimental value for the $K$-meson decay parameter $\epsilon $
we get the value of $\lambda _{LR}^W\leq 10^{-6}$.
\end{center}

The other calculation\cite{herc} of the pion-nucleon CP-odd coupling
constant in this model provides the more optimistic value $\lambda
_{LR}^W\sim 4\times 10^{-3}$. This result is directly dependant on the
restriction on the CP-odd parameters of the model $\left| \zeta \sin (\alpha
-\delta _2)\right| \leq 1.7\times 10^{-3}$ obtained from an experiment on $%
^{19}Ne$-decay \cite{nedec}. However, the restriction on these parameters $%
\left| \zeta \sin (\alpha -\delta _2)\right| \leq 3\times 10^{-6}$ obtained
in paper\cite{frere} from the measurements of neutron electric dipole moment 
\cite{edm1,edm2} ($\left| D_n\right| \leq 1.1\times 10^{-25}e\cdot cm$)
leads to the value $\lambda _{LR}^W\sim 10^{-7}.$ It should be noted, that
such a strong restriction on the CP-violating parameters in the left-right
model has been obtained as a result of barring accidental cancellations in
the QCD short-distance coefficient of the exchange diagram for the neutron
dipole moment calculation\cite{frere}. However, these QCD corrections to the
exchange diagram are very sensitive to the calculation approaches and to the
long-distance QCD parameters.

\subsection{Models of class (c).}

The classical example of the CP violation due to complex vacuum expectation
values of Higgs bosons is the Weinberg model of spontaneous CP- violation%
\cite{wein}. For this model it is convenient to estimate the parameter $%
\lambda $ as the ratio of the CP-odd $g_{CP}$ to P-odd $g_P$ pion-nucleon
coupling constants. We use the value of the $g_P\simeq 1.6\times 10^{-7}$%
(see ref.\cite{dub}). For CP violation from charged Higgs bosons exchange
the effective CP-violating Lagrangian \cite{ad}

\begin{equation}
\label{lagw}{\cal L}_{CP}=2{\it Im}\{A\}m_um_d\cos {^2\theta _c}\times [( 
\overline{d}u)(\overline{u}i\gamma _5d)+(\overline{d}i\gamma _5u)(\overline{u%
}d)]. 
\end{equation}
leads \cite{bgcon} to

\begin{eqnarray}  
g^{ch}_{CP} = \langle {n\pi^+}|{\cal L}_{CP}|{p} \rangle  \nonumber \\
\simeq {\it Im}\{ A\} {{m_u-  
m_d}\over{m_u+m_d}}&m^2_{\pi}f_{\pi}{{3M_{\Lambda}+M_{\Sigma}  
-2M_p}\over{m_s-m_u}} \cos{^2\theta_c} . \label{gwch}  
 \end{eqnarray}   

Here $m_\pi $ and $f_\pi $ are pion mass and the decay constant, $m_q$ is
the quark mass, $M_p$ and $M_{\Lambda ,\Sigma }$ are proton and hyperon
masses, and $\theta _c$ is the Cabibbo angle and $A$ is the propagator of
changed Higgs bosons.

For neutral Higgs bosons, the corresponding constant $g^0_{CP}$ can be
written as\cite{bgcon}

\begin{equation}
\label{gs}g_{CP}^0=g_s<\sigma H><H|\pi >m_\pi ^2, 
\end{equation}
where $g_s$ is a scalar ($\sigma $) Higgs boson-nucleon coupling constant, $%
<H|\pi >$ is the pseudoscalar ($H$) Higgs boson-pion mixing amplitude, and $%
<\sigma H>$ is the neutral Higgs boson propagator.

Due to the anomaly in the energy-momentum tensor, the vertex $g_s$ is
proportional to the nucleon mass \cite{svz} $M$:

\begin{equation}
g_s = -8/29 (M/v), 
\end{equation}

where $v=(G\sqrt{2})^{-1/2}$. If the ''up''-quarks and ''down''-quarks
obtain their masses from different Higgs fields, then the expression for the
pseudo-scalar Higgs boson and meson coupling is similar to the corresponding
expression in the axion theory \cite{ans,anur}:

\begin{equation}
<H\vert \pi> = {\frac{f_{\pi}}{{2\sqrt{2}v}}}\left[ x\left(1-N{\frac{{1-z}}{{%
1+z}}}\right) - {\frac{1}{{x}}}\left(1+N{\frac{{1-z}}{{1+z}}}\right)
\right]. 
\end{equation}

Here $z=m_u/m_d$ , $N$ is the number of quark generations, $x$ is the ratio
of the VEV's corresponding to ''up''-and ''down''-quark masses. Then, using
these expressions with $x=1$ and $N=3$, we can obtain \cite{bgcon}

\begin{equation}
\label{gses}g^0_{CP} = {\frac{{12\sqrt{2}}}{{29}}}{\frac{{<\sigma H>}}{{v^2}}%
}Mf_{\pi}m^2_{\pi}{\frac{{m_d-m_u}}{{m_d+m_u}}} . 
\end{equation}

The old estimations \cite{ad} of the propagator ${\it Im}\{A\}\simeq G\times
0.25GeV^{-2}$ and an assumption that $<\sigma H>/v^2\simeq {\it Im}\{A\}$
gave very large parameters $\lambda $: for CP violation from charged Higgs
bosons exchange $\lambda _H^{ch}\sim 10^{-4}$and for CP violation from
neutral Higgs bosons exchange $\lambda _H^0\sim 10^{-1}$. The parameter $%
\lambda _H^{ch}$ is proportional to $m_H^{-2}$ (where $m_H$ is a mass of
charged Higgs boson) and the given value corresponds to a very small mass of
Higgs boson $m_H\sim 2GeV$ . The given value for the parameter $\lambda _H^0$
is in a contradiction\cite{ans} with the restriction on neutron electric
dipol moment. The up-to-date estimations give $\lambda _H^{ch}\leq 2\times
10^{-6}$ and $\lambda _H^0\leq 10^{-3}$.

\subsection{Models of class (d).}

The CP violation due to neutral Higgs boson exchange can be described in
terms of pure gluonic operators. From this point of view the contribution
from neutral Higgs bosons exchange discussed in the previous section
corresponds to an effective dimension eight four gluonic operator $\sim GGG
\widetilde{G}$, where $\tilde G_{\mu \nu }^a=\frac 12\epsilon _{\mu \nu \rho
\sigma }G_{\rho \sigma }^a$. Indeed, the $GG$ part of the operator
corresponds to the scalar boson nucleon coupling and the $G\widetilde{G}$
part corresponds to the pseudo scalar one in eq.(\ref{gs}). This operator is
a dominant for an exchange of light Higgs bosons. In the case of a heavy
Higgs bosons exchange ( $m_H\geq 100GeV$) the dominant operator is the
dimension six pure gluonic CP- odd Weinberg operator $GG\widetilde{G}$ \cite
{wgl}. The CP-odd Lagrangian for the Weinberg operator can be written as 
\cite{u}

\begin{equation}
\label{lag}{\cal L}=\chi \times \eta _{QCD}\times \tilde O, 
\end{equation}

\begin{equation}
\label{op}\tilde O={\frac{{g_s^3}}{{(4\pi )^2}}}f_{abc}G_{\mu \nu }^aG_{\nu
\rho }^b\tilde G_{\rho \mu }^c. 
\end{equation}
Here $g_s$ is the QCD gauge coupling constant; $\eta _{QCD}$ is the
radiative QCD correction parameter; and $\chi $ is the dimension coefficient
which can be calculated for specific CP- violating models. The parameter $%
\lambda $ for that operator $\lambda _G$ is calculated as the ration of
CP-odd ($g_{CP}$) to P-odd ($g_P$) pseudo scalar meson - nucleon coupling
constants. According to eq.(\ref{lag}), one can write

\begin{equation}
\label{ccp}g_{CP}=\chi \times \eta _{QCD}\times M,
\end{equation}
where $M=\langle {Np}|\tilde O|{N}\rangle $ is the nucleon-pseudo scalar
meson matrix element for the $\tilde O$ operator. Accepting the estimation%
\cite{g3g} of this hadronic matrix element $M\sim 0.2GeV$ and the value of
P-odd pion-nucleon coupling constant\cite{dub} $g_P\simeq 1.6\times 10^{-7}$%
, we get the parameter $\lambda _{G\text{ }}$as:

\begin{equation}
\lambda _G\simeq 10^6m_p^2\chi \ \eta _{QCD}, 
\end{equation}
where $m_p$ is the proton mass.

Using the calculations of the coefficient $\chi $ in paper\cite{u}, one can
obtain the values for the parameter $\lambda _G$ in different models of
CP-violation. In the case of CP violation due to Higgs bosons exchange and
under an assumption about the reasonable scale of the Higgs boson and $t$%
-quark masses ($2\cdot m_H\sim m_t\sim 200GeV$), one obtains\cite{g3g}:

\begin{equation}
\lambda _{Higgs}\sim (0.2-1.0)\times 10^{-2}{\sl Im}Z, 
\end{equation}

where $Z$ is Higgs mixing parameter. Taking into account the bound on the
parameter ${\sl Im}Z\leq 0.03$ (which was obtained\cite{u} from the
experimental limit on the NEDM) one has

\begin{equation}
\lambda _{Higgs}\leq 3\times 10^{-4}. 
\end{equation}

It should be noted, that the left-right model also can lead to the CP-odd
three gluonic operator due to complex CP-odd mixing of left and right
bosons. In this case, for the model with equal gauge coupling constants for
the right and left bosons and when $M_R\gg M_L$ (where $M_{R(L)}$ is a mass
of the right (left) boson) one obtains (see, also ref.\cite{u,ch})

\begin{equation}
\lambda _{LR}\sim 0.1 \sin {\alpha }\sin {\xi }. 
\end{equation}

Here $\xi $ and $\alpha $ are a CP-even and a CP-odd mixing angles of the
left and right bosons. The result for the parameter $\lambda $ is dependent
on the restriction on the parameters ($\sin {\alpha }\sin {\xi }$)${\ }$and${%
\ }$ leads to the following values : $\lambda _{LR}\leq 2\cdot 10^{-4}$ or $%
\lambda _{LR}\leq 4\cdot 10^{-7}$ (see discussion at the end of section
''Model of Class (b)'').

It should be noted, that the contribution from the Kobayashi - Maskawa model
through the Weinberg three gluonic operator is almost negligible\cite{u}.
This is a consequence of the three loop contribution to the coefficient $%
\chi $ which leads to the suppression factor $\sim (m_b/M_W)^4$.

In the same way, one can calculate the value of $\lambda $ for the $\theta $%
-term in QCD using the CP-odd pion-nucleon coupling constant which was
obtained in paper\cite{theta}. The experimental restriction on the neutron
electric dipole moment leads to the following restriction on the parameter $%
\lambda _\theta $:

\begin{equation}
\lambda _\theta \leq 5\times 10^{-5}. 
\end{equation}

Taking into account that an accuracy for the existing calculations of the
P-odd and CP-even nucleon coupling constants and the P-odd and CP-odd ones
is about of one order of magnitude, the expected accuracy for the parameter $%
\lambda $ is approximately at the same level. It means that a model with a
rather large value of the parameter $\lambda $ (for example, $\lambda \sim
10^{-2}$) might exist in each class of CP-violating models. It should be
noted, that in spite of this conclusion the well known models of CP
violation give a rather small value of the $\lambda \leq 10^{-3}$ (see the
above discussions).

The given calculations of the parameter $\lambda $  lead to the conclusion
that in every class of CP-violating models the parameter $\lambda $ might be
large enough to be measured in the neutron scattering experiment. To improve
the current restrictions on the  $\lambda $ the further theoretical
investigations are needed. 

\section{The in-medium behavior of CP-odd coupling constants}

The CP-violating coupling constants calculated in vacuum provide correct
results in nuclear matter for almost all models because we are interested in
the parameter $\lambda ,$ which is the ratio of the CP-odd to the P-odd
nucleon coupling constants. If the origin of CP violation is not related to
the strong interaction, this ratio for nuclear matter must be the same as
for the vacuum free particle interaction. From this point of view, the model
of CP violation due to the $\theta $-term in QCD Lagrangian is a rather
special case because the mechanism of CP violation is related to the
properties of the strong interaction. Therefore, the relative value of
CP-odd effects in nuclear matter may be changed in comparison to the vacuum
case. This problem has been considered in paper \cite{gvac}. Since the
measure of CP violation in vacuum due to the $\theta $-term in QCD is \cite
{svzm}: 
\begin{equation}
\label{vaccp}\kappa _{vac}={\frac{\langle {\frac{\alpha _s}{{\pi }}}G\tilde
G\rangle _{vac}}{{\langle {\frac{\alpha _s}{{\pi }}}GG\rangle _{vac}}},} 
\end{equation}
the measure for the CP violation in nuclear matter is

\begin{equation}
\label{mcp}\kappa _\rho ={\frac{\langle {\frac{\alpha _s}{{\pi }}}G\tilde
G\rangle _\rho }{{\langle {\frac{\alpha _s}{{\pi }}}GG\rangle _\rho }}.}
\end{equation}
Here ${{\langle {\frac{\alpha _s}{{\pi }}}GG\rangle _{vac}}}$ is a vacuum
gluon condensate,  ${\langle {\frac{\alpha _s}{{\pi }}}G\tilde G\rangle
_{vac}}$ is a condensate of CP-odd gluonic operator ${{\frac{\alpha _s}{{\pi 
}}}G\tilde G}$ in the vacuum and  $\langle \ \rangle _\rho $ are the
corresponding in-medium condensates. From this expression one can see that a
renormalization of CP-odd effects in nuclear matter is defined by the
renormalization of the CP-odd operator ${\langle {\frac{\alpha _s}{{\pi }}}%
G\tilde G\rangle _\rho }$, ${\ }$because the gluon condensate is just
slightly changed in nuclear matter at the saturation density\cite{qgc}.

The nuclear density dependence for the operator ${\langle {\frac{\alpha _s}{{%
\pi }}}G\tilde G\rangle \ }$ \cite{gvac} 
\begin{equation}
\label{tro}{\frac{\langle {\frac{\alpha _s}{{\pi }}}G\tilde G\rangle _\rho }{%
{\langle {\frac{\alpha _s}{{\pi }}}G\tilde G\rangle _{vac}}}}\simeq 1+{\frac
\rho {{\langle \overline{q}q\rangle _{vac}}}}{\frac{\sigma _N}{{(m_u+m_d)}}}{%
\ \frac{(m_u+m_d)^2}{{4m_um_d}}}
\end{equation}
is  the same one as for the quark condensate \cite{qgc}

\begin{equation}
\label{qren}{\frac{\langle \overline{q}q\rangle _\rho }{{\langle \overline{q}%
q\rangle _{vac}}}}\simeq 1+{\frac \rho {{\langle \overline{q}q\rangle _{vac}}%
}}{\frac{\sigma _N}{{(m_u+m_d)}}.}
\end{equation}
Here $\rho $ is the medium (nuclear matter) density; $\sigma _N$ is the
nucleon $\sigma $-term; $m_{u,d}$ are current masses of $u,d$-quarks. The
additional multiplier in eq.(\ref{tro}) is not significant: $%
(m_u+m_d)^2/(4m_um_d)=1.08$. Taking into account that the quark condensate
may be reduced in its value by about ($25\%-50\%$) at the nuclear saturation
density\cite{qgc}, one can conclude that the CP-odd interaction due to $%
\theta $-term has the in-medium reduction.

The CP-odd coupling constant $g_{CP}$ is proportional to the measure of CP
violation $\kappa $$_\rho $ and, consequently, has the same density
dependence as the quark condensates (by neglecting the gluon condensate
density dependence). From the other hand, the P-odd coupling constant $g_P$
is, also, proportional to the quark condensate value (see, e.g. ref.\cite
{dub}). Therefore, the parameter $\lambda =g_{CP}/g_P$ has a negligible
density dependence.

It should be emphasized that the approximation used for description of the
quark condensates behavior in nuclear matter\cite{qgc} has an accuracy of
about $10\%$ up to the nuclear saturation density. Therefore, the above
conclusion is valid to the same accuracy.

\section{ CP-odd nucleon potential}

For the estimations of nuclear matrix elements in section 2 we used simple
one particle nuclear potentials. To calculate the parameter $\lambda $ for
real experiments it is desirable to use CP-odd one boson exchange
potentials. Since the $\pi $-meson contribution is dominant for the CP-odd
one boson exchange interactions\cite{ghm}, the following one meson CP-odd
potential\cite{hax,herc}

\begin{eqnarray}
V_{CP}^\pi = &&-{\frac{m_\pi ^2}{8\pi m_N}%
}g_{\pi NN}\cdot \vec r{\frac{e^{-m_\pi r}}{m_\pi r^2}}[1+{\frac
1{m_\pi r}}] \nonumber \\ 
&&\times [ \bar g_{\pi NN}^{(0)}\cdot
(\vec \tau_1\cdot \vec \tau _2) \cdot (\vec \sigma_1-\vec \sigma_2)\nonumber \\
&&+{{\bar g_{\pi NN}^{(1)}}\over{2}}\cdot [(\tau_{1z}+\tau _{2z}) \cdot
(\vec \sigma_1-\vec \sigma_2)+(\tau_{1z}-\tau_{2z})\cdot (\vec \sigma_1+\vec
\sigma_2)]\nonumber \\
 && + \bar g_{\pi NN}^{(2)} \cdot (3\tau _{1z}\tau_{2z}-
\vec \tau _1 \cdot \vec \tau_2) \cdot 
(\vec \sigma_1-\vec \sigma_2) ]  \label{potent}
\end{eqnarray}

is a good approximation for the description of CP violation in nuclei. Here $%
g$ and $\bar g^{(T)}$ are strong CP-even and weak CP-odd pion-nucleon
coupling constants; $T=0,1,2$ correspond to isoscalar, isovector and
isotensor interactions, respectively;  ${\bf \vec \sigma }$ and ${\bf \vec
\tau }$ are the spin and isospin of the nucleon.

Therefore, to calculate CP-violating effects in nuclei in terms of the one
boson nucleon interactions one can use only $\pi $-meson nucleon CP-odd
parameters $\bar g^{(T)}$. The simple structure of the CP-odd nucleon
potential gives the opportunity to calculate CP-violating effects and leads
to the simple parameterization of all CP-odd effects in nuclei using only $%
\pi $-meson parameters. This fact provides the opportunity to test different
models of CP violation with a good accuracy in the framework of the given
parameterization.

It should be noted, that different models of CP violation usually give
contributions not to all three parameters $\bar g^{(T)}$ , but rather to
some of them. Therefore, the real potential for the particular model is
usually even much simple then the one given in eq.(\ref{potent}).

\section{Conclusions}

The study of T-violating correlations in neutron scattering leads to a
unique opportunity to search for CP violation because of the large
enhancement of experimental CP-odd effects in the vicinity of p-wave
resonances and the possibility to calculate CP-odd nuclear effects starting
from the original model of CP violation at the quark-gluon level. The
estimated CP-violating effects for some models show that each class of
CP-violating models can give a measurable effect for the neutron scattering
experiments. Even if CP violation will not be detected in the neutron
scattering experiments, the obtained experimental data could give the
unambiguous restrictions on many models of CP violation. Since the nucleon
CP-odd potential has the main contribution from one $\pi $-meson exchange it
is possible to obtain a direct relation between CP-odd nuclear effects and
the value of the neutron electric dipole moment if the one meson loop gives
the main contribution\cite{theta,hm,chem}.

Therefore, the results which can be obtained from the neutron scattering
experiments might be of the same accuracy and importance as, for example,
the results to be expected from the currently fashionable B-meson physics or
from a measurement of the neutron electric dipole moment.

\end{document}